\documentclass[english]{article}
\usepackage[T1]{fontenc}
\usepackage[latin9]{inputenc}
\usepackage{float}
\usepackage{amssymb}
\usepackage{graphicx}
\usepackage{natbib}
\RequirePackage{amsthm}
\RequirePackage[cmex10]{amsmath}
\makeatletter

\providecommand{\tabularnewline}{\\}
\floatstyle{ruled}
\newfloat{algorithm}{tbp}{loa}
\providecommand{\algorithmname}{Algorithm}
\floatname{algorithm}{\protect\algorithmname}

\theoremstyle{plain}

\@ifundefined{showcaptionsetup}{}{%
 \PassOptionsToPackage{caption=false}{subfig}}
\usepackage{subfig}
\makeatother

\usepackage{babel}
\begin{document}
\title{Approximation-Robust Inference in Dynamic Discrete Choice}
\author{Ben Deaner}
\maketitle
\begin{abstract}
Estimation and inference in dynamic discrete choice models often relies
on approximation to lower the computational burden of dynamic programming.
Unfortunately, the use of approximation can impart substantial bias
in estimation and results in invalid confidence sets. We present a
method for set estimation and inference that explicitly accounts for
the use of approximation and is thus valid regardless of the approximation
error. We show how one can account for the error from approximation
at low computational cost. Our methodology allows researchers to assess
the estimation error due to the use of approximation and thus more
effectively manage the trade-off between bias and computational expedience.
We provide simulation evidence to demonstrate the practicality of
our approach.
\end{abstract}
Models of dynamic decision making have been applied to a wide range
of important problems in empirical economics (see \cite{Keane2011}
for a survey). In these models agents face a dynamic decision problem
that can often be solved only numerically and at substantial computational
cost. Estimation of these models generally requires that
the empirical practitioner solve the agents' dynamic programming problem for
each candidate set of parameter values as the likelihood is maximized
or GMM objective minimized.\footnote{Exceptions include the constrained optimization approach of \cite{Su}
and the conditional choice probability approaches of \cite{Hotz1993}
and \cite{Aguirregabiria2007}.} For estimation to be computationally feasible, the researcher may
need to use approximate dynamic programming methods. The most common
approach in empirical economics is approximation of the `value function' or of related functions like the expected value function or the choice-specific expected value
function. The function is iteratively
interpolated between a small subset of states using say, polynomials,
splines, or kernel regression.

Approximation of the solution to the dynamic programming problem has
a long history both in operations research and in econometrics. \cite{Bellman1963l}
suggests a method based on polynomial approximation of the expected
value function. Approximate dynamic programming methods were introduced
to empirical economics by \cite{Keane} who developed an interpolation
method designed for use in finite-horizon dynamic discrete choice
models. More recent work includes that of \cite{Norets2012} who approximates
the expected value function using artificial neural networks in the
dynamic discrete choice setting.

In many cases value function approximation occurs implicitly. When
the state space is continuous some kind of interpolation of the value
function is unavoidable, and in many cases models with a discrete
state-space implicitly approximate a true model with a continuous
state-space.

Unfortunately, the use of approximation leads to inconsistent estimates,
and confidence intervals that do not account for the approximation
error are not asymptotically valid. One employs approximation methods
hoping that the resulting bias is not of too great a magnitude. Without
explicit bounds on the approximation error it is unclear how close
the approximated value function is to the true value function and
it is also unclear how sensitive the estimates are to the approximation
error.

Monte Carlo simulation can be used to assess the biases due to approximation
(see \cite{Keane}) but simulation studies are inherently limited.
Firstly, the findings of simulation studies apply only to the specific
model that is simulated. Secondly, one must be able to exactly solve
the dynamic programming problem in order to simulate the data, which
means that simulation studies can only be carried out on relatively
simple models without a very large state space.

Ideally estimation and inference would explicitly account for the
use of approximation. We show how this can be achieved. Using the
contraction mapping property of the Bellman operator as well as some
other features of the problem, we derive bounds on the differences
between choices of the choice-specific expected value function. These
bounds can be calculated at relatively little computational cost.
Using these bounds one can attain a set estimate that must contain
the infeasible estimate that could be evaluated if the researcher
were able to fully solve the dynamic programming problem. The bounds
can be used to achieve confidence sets that are valid regardless of
the closeness of the approximation.

Bounds on the approximation error in approximate dynamic programming
are considered in for example \cite{Arcidiacono2013}, but these bounds
are generally on the value function, expected value function, or choice-specific expected value function itself. The value function enters
the likelihood only through the differences between the the expected
values at different choices. We directly bound the latter and then, using these bounds, derive error bounds on the likelihood.

We develop set
estimates and confidence sets that are asymptotically valid regardless
of the distance between the exact value function. Our methodology is valid regardless of the method used to approximate the value function.

In short, we provide a methodology that allows researchers to conduct
asymptotically valid inference and consistent set estimation, regardless
of the accuracy of their value function approximation. Of course,
if the approximation is poor then the resulting set estimates and
confidence sets will be large, but if the approximation is close they
will be small.

Our methodology allows a researcher to directly assess the closeness
of their approximation, and balance the trade-off between ease of
computation and accuracy of estimation. If the researcher observes
that the set estimates and robust confidence intervals are very large compared to standard non-robust confidence sets,
then they can decrease the coarseness of their discretization of the
state space, increase the order of the approximating polynomial, etc.
The tools presented in this paper also apply to novel methods of approximate
dynamic programming like Q-learning. Economic researchers might be
reluctant to try less well-established methods because they fear that
the strength of the approximation may be poor. Our methodology can
allay these fears because it allows researchers to explicitly assess
the biases due to approximation error and perform estimation and inference
that is robust to this type of error.

The paper is organized as follows: Section 1 describes the general
model in which the results apply and provides background to the problem.
Section 2 provides the derivation of bounds on the value function
differences. Section 3 explains how the bounds can be used for estimation
and inference that is robust to approximation.
Section 4 provides Monte Carlo evidence of the efficacy of the method,
and Section 5 concludes.

\section{Background}

\subsection*{Modeling Dynamic Discrete Choices}

Consider an agent who at a each time $t$ is at some state $S_{t}\in\mathcal{S}$
and makes choice $D_{t}\in\mathcal{D}$. The `choice set' $\mathcal{D}$
is assumed to be discrete, but we make no such assumption about the
`state space' $\mathcal{S}$. The state evolves according to a first-order Markov process in which the probability distribution of next
period's state may depend on both the current state $S_{t}$ and choice
$D_{t}$, and is otherwise unrelated to the history of states and
choices. The static utility of the agent in each period is given by
$u(S_{t},D_{t})+\epsilon_{t}(D_{t})$ where $\epsilon_{t}$ is an
additive iid random function from $\mathcal{D}$ to $\mathbb{R}$.
We refer to $\epsilon_{t}$ as `shocks' to the utility because they
are not known to the agent prior to period $t$. We assume throughout
that $E[\max_{d\in\mathcal{D}}\epsilon_{t}(d)]<\infty$. We assume
the deterministic part of the utility $u$ is bounded, i.e., there
exists a scalar $b<\infty$ so that $|u(s,d)|\leq b$ for all $s\in\mathcal{S}$
and $d\in\mathcal{D}$.

The agent sets $D_{t}=d(S_{t},\epsilon_{t})$, where $d(\cdot,\cdot)$
is a static decision rule that maximizes the expected discounted (by
discount factor $\beta$) sum of utilities:
\[
d\in\underset{d:S\times E\to D}{\arg\sup}E\bigg[\sum_{t=1}^{\infty}\beta^{t}u\big(S_{t},d(S_{t},\epsilon_{t})\big)+\epsilon_{t}\big(d(S_{t},\epsilon_{t})\big)\bigg]
\]
There are potentially two sources of uncertainty in the agent's problem.
Firstly, $\epsilon_{t}$ is a random function. Secondly, the transition
between states may be stochastic. We denote by $F_{s,d}$ the probability
distribution of the random variable $S_{t+1}$ for an agent who at
time $t$ is at state $S_{t}=s$ and chooses $D_{t}=d$. $F_{s,d}$
is referred to as the `transition distribution' at $(s,d)$ and we
refer to $F$, which maps from $(s,d)$ to $F_{s,d}$ as the `transition
rule'. That the state in the next period depends on the decision in
the current period is what makes the agent's problem `dynamic': the
decision today effects not only today's utility, but the utility tomorrow
because it partially determines tomorrow's state.

If the dynamic optimization problem is sufficiently regular, then
the decision rule $d(s,\epsilon)$ is the unique solution to the following
static optimization problem:
\[
d(s,\epsilon)=\underset{d\in\mathcal{D}}{\arg\max}[u(s,d)+\epsilon(d)+\beta V(s,d)]
\]

Where $V:\mathcal{S}\times\mathcal{D}\to\mathbb{R}$ is the unique
fixed point of the non-linear operator $T:\mathbb{R}^{(\mathcal{S}\times\mathcal{D})}\to\mathbb{R}^{(\mathcal{S}\times\mathcal{D})}$
which is defined below:

\begin{equation}
T[V](s,d)=E\big[E[\max_{d'}\,u(S_{t+1},d)+\beta V(S_{t+1},d)+\epsilon_{t+1}(d)|S_{t+1}]\big|S_{t}=s,D_{t}=d\big]\label{eq:bell}
\end{equation}

And so $V$ solves the non-linear operator equation:
\begin{equation}
V=T[V]\label{eq:belleq}
\end{equation}

This characterization of the optimal decision rule is due to the work
of Bellman (see \cite{Bellman1957}), and is therefore known as `Bellman's
principle of optimality'. We refer to $V$ as the `choice-specific
expected value function' or just as the `value function' for short.
Note that in the literature the `value function' usually refers to
instead to the quantity $\max_{d}u(s,d)+\epsilon(d)+\beta V(s,d)$
which is a function of both $s$ and $\epsilon$. $\beta V(s,d)$
is the expected discounted future returns from time $t+1$ onwards
for an agent who, at time $t$, is at state $s$ and makes choice
$d$, and acts optimally thereafter. We refer to (\ref{eq:belleq})
as the `Bellman equation' and $T$ as the `Bellman operator' although
these terms are often used slightly differently (see \cite{Keane2011}).

\subsection*{Point Estimation}

To create a statistical model within the dynamic discrete choice framework
let us introduce a vector of parameters of interest $\theta$. $\theta$
may parameterize the utility function, the distribution of $\epsilon_{t}$,
and the state transition rule $F$. For simplicity we focus on the
case in which $\theta$ parameterizes the utility and distribution
of $\epsilon_{t}$. Both the utility function and distribution of
$\epsilon_{t}$ are treated as known up to the parameters $\theta$.
To emphasize the dependence of the utility function on $\theta$ we
let $u(s,d;\theta)$ be the deterministic part of the utility at state
$s$ and from choice $d$ under the parameters $\theta$.

Note that the Bellman operator (\ref{eq:bell}) depends on the utility
function, thus the solution $V$ to the Bellman equation (\ref{eq:belleq})
depends on the parameters $\theta$ and discount factor $\beta$.
In addition, the Bellman operator depends on the transition rule $F$
and therefore so does the solution $V$. To make clear this dependence
we write $V_{(\theta,\beta,F)}(s,d)$ to denote the choice-specific
expected value function at choice $d$ and state $s$ when the utility parameters
are equal to $\theta$, the discount factor equals $\beta$, and the
transition rule is $F$.

It will be convenient to define the differences in shocks, utilities
and choice-specific values between pairs of choices $d$ and $d'$:
\begin{align*}
\Delta_{d,d'}\epsilon_{t} & =\epsilon_{t}(d)-\epsilon_{t}(d')\\
\Delta_{d,d'}u(s;\theta) & =u(s,d;\theta)-u(s,d';\theta)\\
\Delta V_{(\theta,\beta,F)}(s,d,d') & =V_{(\theta,\beta,F)}(s,d)-V_{(\theta,\beta,F)}(s,d')
\end{align*}
 Given the model in the previous section, the probability that an
agent at state $S_{t}=s$ makes decision $D_{t}=d$ is given by:

\begin{equation}
L_{(\theta,\beta;F)}(s,d)=P_{\theta}\bigg[\Delta_{d,d'}\epsilon_{t}\geq\Delta_{d',d}u(s;\theta)+\beta\Delta V_{(\theta,\beta,F)}(s,d',d),\forall d'\in\mathcal{D}\bigg]\label{eq:likely}
\end{equation}

Where $P_{\theta}$ is the probability given the distribution of $\epsilon_{t}$
which possibly depends on parameters $\theta$. If the transition
rule $F$ were known then $L_{(\theta,\beta;F)}(S_{t},D_{t})$ is
the likelihood of the observation $(S_{t},D_{t})$ at time $t$. If
$F$ is unknown then it is a nuisance parameter. $F$ can be estimated
directly from the data in a first stage either parametrically or nonparametrically.
Let $\hat{F}$ be some estimate of $F$ that is evaluated in a first
stage. With $F$ replaced by the estimate $\hat{F}$, the partial
log-likelihood of the data $\{S_{t},D_{t}\}_{t=1}^{T}$ evaluated
at parameters $\theta$ and $\beta$ is then given by:
\[
\mathcal{L}_{\{S_{t},D_{t}\}_{t=1}^{T}}(\theta,\beta;\hat{F})=\sum_{t=1}^{T}\ln L_{(\theta,\beta;\hat{F})}(S_{t},D_{t})
\]

An estimate of the parameters $\theta$ and $\beta$ is then given
by:

\[
(\hat{\theta},\hat{\beta})=\arg\max_{\theta,\beta}\mathcal{L}_{\{S_{t},D_{t}\}_{t=1}^{\infty}}(\theta;\beta,\hat{F})
\]

Identification of the discount factor can be difficult (see \cite{Rust1987},
\cite{Magnac2002}, \cite{Abbring}), and so the discount factor may
be calibrated in some applications. In this case one can simply plug-in
the assumed value for $\beta$ and maximize the log-likelihood only
over the argument $\theta$. The maximization problem above seldom
admits a closed-form solution, and so numerical optimization methods
are required. Thus, for a number of candidate values of $\theta$
and possibly $\beta$ one must evaluate $\mathcal{L}_{\{S_{t},D_{t}\}_{t=1}^{T}}(\theta,\beta;\hat{F})$
and/or its derivatives. In order to calculate $\mathcal{L}_{\{S_{t},D_{t}\}_{t=1}^{T}}(\theta,\beta;F)$
one must evaluate the value function differences $\Delta V_{(\theta,\beta,F)}(s,d',d)$
at all choices in $\mathcal{D}$ and at all states observed in the
data. This procedure is known as the `nested fixed point' or NFXP
algorithm because nested within each step of a numerical maximum likelihood
procedure one applies an algorithm to find a fixed point of the Bellman
operator (\cite{Wolpin1984}, \cite{Rust1987}).

The standard method for evaluating $V$ at all realized states up
to close approximation is `Bellman iteration'. To apply the method
one begins with some guess of the solution $V_{(\theta,\beta,F)}(\cdot,\cdot)$
and then iteratively applies the Bellman operator until there is little
change between iterations. Let us be more precise. Let $T_{(\theta,\beta,F)}$
be the Bellman operator (defined in (\ref{eq:bell})) when the parameters
are set to values $\theta$, $\beta$ and $F$. For any $n\in\mathbb{N}$
and $\tilde{V}$, let $T_{(\theta,\beta,F)}^{n}[\tilde{V}]=\big[T_{(\theta,\beta,F)}^{n-1}[\tilde{V}]\big]$
and let $T_{(\theta,\beta,F)}^{1}[\tilde{V}]=T_{(\theta,\beta,F)}[\tilde{V}]$.
Then for some initial guess $\tilde{V}_{0}$ one approximates $V_{(\theta,\beta,F)}(\cdot,\cdot)$
by $T_{(\theta,\beta,F)}^{n}[\tilde{V}_{0}](s,d)$ where $n$ is large
enough that $T_{(\theta,\beta,F)}^{n}[\tilde{V}_{0}]$ and $T_{(\theta,\beta,F)}^{n-1}[\tilde{V}_{0}]$
sufficiently close in some sense. Because $T_{(\theta,\beta,F)}$
is a contraction mapping, one can show that for any bounded function
$\tilde{V}_{0}$, $T_{(\theta,\beta,F)}^{n}[\tilde{V}_{0}]$ converges
in the supremum norm to the unique fixed point of $T_{(\theta,\beta,F)}$
as $n\to\infty$. \footnote{In practice it may be more computationally efficient to combine this
procedure with a faster but less stable method like Newton-Kontorovich
(\cite{Rust1987}).}

Unfortunately, each iteration of the fixed point procedure above requires
that one evaluate the right hand side of (\ref{eq:bell}) at every state
$s$ and choice $d$, not just those states and decisions observed
in the data.\footnote{Strictly speaking one can restrict the state-space to only those states
for which there exists a decision rule $d(\cdot,\cdot)$ under which
the state is reached with positive probability starting at the states
observed in the data.} If the state space is discrete but large then this can be computationally
burdensome and if the state space is not discrete and finite it is
generally impossible. This same problem applies for many alternative
methods of finding a fixed point to the Bellman equation like Newton-Kantorovich.

An alternative to the NXFP algorithm, Mathematical Program with Equilibrium
Constraints (MPEC) avoids the need for Bellman iteration (\cite{Su}).
MPEC treats the estimation problem as a constrained optimization problem
in which both $\theta$ and (possibly) $\beta$ and the value function
itself are parameters and the the Bellman equation (\ref{eq:belleq})
is treated as a constraint. Thus in MPEC the number of scalar constraints
is proportional to the cardinality of the state space. As such, when
the state space is large MPEC may still be computationally burdensome.

In many cases, the RHS of (\ref{eq:bell}) cannot be evaluated exactly
even at a single state and choice. The expectation over the future
states on the RHS of (\ref{eq:bell}) generally does not have an analytical
form and so must be evaluated by numerical methods. The inner expectation
on the RHS of (\ref{eq:bell}) may also be analytically intractable.
For instance, this is the case when 
$\epsilon_{t}$ (understood as a vector with each entry corresponding to $\epsilon _t (d)$  for some $d\in \mathcal{D}$) is multivariate normal with a non-diagonal covariance
matrix. This further compounds the computational burden of applying
the Bellman operator at a large number of states.

To avoid the need to evaluate the RHS of the Bellman equation at a
large and possibly infinite number of states, some kind of approximation
is needed. For example, one may apply the Bellman operator at only
some subset of states and then for other states approximate the value
function as needed by interpolation. Let us denote by $\tilde{T}_{(\theta,\beta,F)}$
an approximate Bellman operator so that $\tilde{T}_{(\theta,\beta,F)}[\tilde{V}](s,d)=T_{(\theta,\beta,F)}[\tilde{V}](s,d)$
for all $s$ in some small, finite subset $\tilde{\mathcal{S}}\subset\mathcal{S}$
and all choices $d\in\mathcal{D}$ and for all other $s\in\mathcal{S}$
and $d\in\mathcal{D}$ takes some value given by interpolation. We
refer to $\tilde{\mathcal{S}}$ as a `discrete grid'. It is important
to note that in general, $\tilde{T}$ is not a contraction. The term
`interpolation' is used here as a catch-all for a range of different
approaches that include fitting a parametric function by least squares,
kernel methods and modern machine-learning regression techniques like
random forests.

If one replaces $\Delta_{(\theta,\beta,F)}V(s;d',d)$ in the expression
for the likelihood (\ref{eq:likely}) with an approximation, then one
evaluates the log-likelihood with error. Thus the resulting maximum
likelihood estimates will generally be inconsistent. When applying
approximation methods one hopes that the approximated value function
differences are close to the exact differences $\Delta_{(\theta,\beta,F)}V(s,d',d)$,
that the approximate log-likelihood is then close to the exact log-likelihood,
and that the estimates are then close to the exact maximum likelihood
estimates and the degree of inconsistency is therefore small. The
need to bound the estimation error that results from value function
approximation motivates the method described in the next section.

Finally, we note that a class of methods referred to as Conditional
Choice Probability (CCP) methods side-step the need for dynamic programming
entirely. This approach was introduced by \cite{Hotz1993}, and the
literature contains a number of different methods based on their approach,
for example \cite{Aguirregabiria2007} and \cite{Bajari2007}. These
methods are based on the observation that under the structure of a
given dynamic discrete choice model, the value function is often equal
to an analytic (or easy to numerically compute) function of the choice
probabilities of the agent at each state. Thus in a first stage one
nonparametrically estimates from the data the probability of each
discrete choice at each stage and plugs this in to get an estimate
of the value function. However, these methods have certain drawbacks.
The first-stage nonparametric choice probability estimation contributes
to the variance and bias of the parameter estimates, and so the parameter
estimates are generally inefficient compared to estimates based on
dynamic programming.\footnote{The method of \cite{Aguirregabiria2002}, which lies somewhere between
CCP and NFXP is efficient up to first order.} This is particularly problematic because the choice probabilities
may be very imprecisely estimates at parts of the state space that
are visited with very low probability (for further discussion see
\cite{Aguirregabiria2007}).

Our method can be used regardless of the manner in which the value
function is approximated or estimated and can therefore complement
CCP methods. Firstly, our method allows for valid inference and consistent
set estimation even when the conditional choice probabilities cannot
be consistently estimated. Secondly, our inference method does not
need to account for any first stage nonparametric estimation of the
conditional choice probabilities and as such confidence sets based
on our method could be smaller than those based on CCP point estimates.

\section{Bounding the Approximation Error}

In the previous section we discussed the need for approximation methods
in the evaluation of the value function differences $\Delta_{(\theta,\beta,F)}V$
at the states observed in the data. In this section we describe how
one can bound the difference between the exact value function differences
$\Delta_{(\theta,\beta,F)}V$ and the differences from some approximate
value function $\tilde{V}$. In Section 3 we show how these bounds
can be used to achieve set estimation and inference that is valid
regardless of the approximation error.

For ease of notation let us again suppress the arguments $\theta$,
$\beta$, and $F$ in the value function $V$ and Bellman operator
$T$. Theorem 1 below provides an upper bound on the distance between
the exact value function differences $V(s,d)-V(s,d')$ and the differences
$\tilde{V}(s,d)-\tilde{V}(s,d')$ for some approximate value function
$\tilde{V}$. The theorem refers to a function $b(s,d,d')$ defined
as follows:
\[
b(s,d,d')=\frac{1-\beta[\sup_{s,s'\in S,d,d'\in D}\delta(F_{s,d},F_{s',d'})]+\beta\delta(F_{s,d},F_{s,d'})}{1-\beta[\sup_{s,s'\in S,d,d'\in D}\delta(F_{s,d},F_{s',d'})]}
\]

Where $\delta(F_{s,d},F_{s',d'})$ is the total variation distance
between $F_{s,d}$ and $F_{s',d'}$.

\theoremstyle{plain}
\newtheorem*{Th1}{Theorem 1}
\begin{Th1}
Suppose $u$ is bounded and $E[\max_{d\in\mathcal{D}}\epsilon_{t}(d)]<\infty$.
Let $V$ solve the Bellman equation (\ref{eq:belleq}) for transition
rule $F$ and discount factor $\beta<1$. For any $d,d'\in\mathcal{D}$
and $s\in\mathcal{S}$ and bounded function $\tilde{V}$ on $(\mathcal{S}\times\mathcal{D})$:
\begin{eqnarray*}
 &  & \big|\big(\tilde{V}(s,d)-\tilde{V}(s,d')\big)-\Delta V(s,d,d')\big|\\
 & \leq & b(s,d,d')\sup_{s,s'\in S,d,d'\in D}|[T[\tilde{V}](s',d')-T[\tilde{V}](s,d)]-[\tilde{V}(s',d')-\tilde{V}(s,d)]]|
\end{eqnarray*}
\end{Th1}
The upper bound given in Theorem 1 does not involve the exact value
function $V$. Instead it involves only the discount factor $\beta$,
the transition law $F$, and the approximate value function $\tilde{V}$.
The upper bound is small when the discount factor $\beta$ is small,
when the transition distribution $F_{s,d}$ does not change much between
states and choices, and when $\tilde{V}$ comes close to satisfying
the Bellman equation. Note that in the case where the approximation
is exact, i.e. $\tilde{V}=V$ the upper bound is equal to zero.

The total variation distance $\delta(F_{s,d},F_{s',d'})$, could be
computed analytically or numerically depending on the transition law
$F$ or first-stage estimate thereof. Note that the total variation distance must be bounded above
by $1$ and so:
\begin{align*}
b(s,d,d') & \leq\frac{1}{1-\beta}
\end{align*}

The inequality above allows us to avoid calculating the total variation
distance but results in a looser bound than that given in Theorem
1. In order to evaluate the upper-bound in Theorem 1 one must also
evaluate the supremum over states $s$ and $s'$ and choices $d$
and $d'$ of the following quantity:
\[
|T\tilde{V}(s,d)-T\tilde{V}(s',d')-[\tilde{V}(s,d)-\tilde{V}(s',d')]|
\]
It may be feasible to directly evaluate the supremum of the above,
however if $\mathcal{S}$ is large then evaluating the supremum is
still computationally expensive. If $\mathcal{S}$ is continuous one
could discretize the state space before finding the supremum, this
of course will still result in approximation error, but this error
may be acceptable because it is not compounded by repeated Bellman
iteration. For the case where the supremum cannot be effectively computed
we describe an algorithm that upper bounds the supremum and that can
be run until the exact supremum and upper bound are arbitrarily close.

Define $\bar{b}_{\tilde{V}}$ by:
\[
\bar{b}_{\tilde{V}}=\sup_{s,s'\in S,d\in D}|u(s,d)+\beta\tilde{V}(s,d)-u(s',d)-\beta\tilde{V}(s',d)|
\]

Note that $\bar{b}_{\tilde{V}}$ can be evaluated without applying
the Bellman operator. Further, because $\bar{b}_{\tilde{V}}$ only
involves $u$ and $\tilde{V}$ which are generally parametric functions,
it may be possible to evaluate (or at least upper bound) $\bar{b}_{\tilde{V}}$
analytically.

Theorem 2 below motivates our algorithm.

\theoremstyle{plain}
\newtheorem*{Th2}{Theorem 2}
\begin{Th2}
Let $\mathcal{A}$ be a finite subset of $\mathcal{S}\times\mathcal{D}$,
then:
\begin{eqnarray*}
 &  & \big|\big(\tilde{V}(s,d)-\tilde{V}(s,d')\big)-\Delta V(s,d,d')\big|\\
 & \leq & b(s,d,d')\sup_{s,s'\in S,d,d'\in D}\bigg[\min_{(s^{*},d^{*})\in\mathcal{A}}\{T[\tilde{V}](s^{*},d^{*})+\delta(F_{s,d},F_{s^{*},d^{*}})\bar{b}_{\tilde{V}}\}\\
 &  & -\max_{(s^{*},d^{*})\in\mathcal{A}}\{T[\tilde{V}](s^{*},d^{*})-\delta(F_{s',d'},F_{s^{*},d^{*}})\bar{b}_{\tilde{V}}\}-[\tilde{V}(s,d)-\tilde{V}(s',d')]\bigg]
\end{eqnarray*}
\end{Th2}

The upper bound in Theorem 2 only requires that one apply the Bellman
operator $T$ at those states and choices in the finite set $\mathcal{A}$.
The supremum in the upper bound may have an analytical solution for
some parametric choices of $\tilde{V}$.

If $\mathcal{A}$ is `dense' in a certain sense, and the functions
involved are uniformly continuous, then the bound in Theorem
2 is not very conservative compared to the bound in Theorem 1. We
formalize this in Theorem 3 below. Theorem 2 suggests the following
procedure to upper bound $\sup_{s,s',d,d'}|T[\tilde{V}](s,d)-T[\tilde{V}](s',d')-[\tilde{V}(s,d)-\tilde{V}(s',d')]|$.
The algorithm returns a quantity $\bar{B}$ that is guaranteed (by
Theorem 2) to upper bound $\sup_{s,s',d,d'}|T[\tilde{V}](s,d)-T[\tilde{V}](s',d')-[\tilde{V}(s,d)-\tilde{V}(s',d')]|$
and that is guaranteed to exceed this quantity by less than a desired
tolerance $\tau$.

\begin{algorithm}[H]
\begin{enumerate}
\item Choose some finite set of states and choices $\mathcal{A}\subset\mathcal{S}\times\mathcal{D}$.
Evaluate $T[\tilde{V}](s,d)$ for each $(s,d)\in\mathcal{A}$. Evaluate
$\bar{b}_{\tilde{V}}$.
\item Calculate $\bar{B}$ given by:
\begin{align*}
\bar{B}= & \sup_{s,s'\in S,d,d'\in D}\bigg[\min_{(s^{*},d^{*})\in\mathcal{A}}\{T[\tilde{V}](s^{*},d^{*})+\delta(F_{s,d},F_{s^{*},d^{*}})\bar{b}_{\tilde{V}}\}\\
 & -\max_{(s^{*},d^{*})\in\mathcal{A}}\{T[\tilde{V}](s^{*},d^{*})-\delta(F_{s',d'},F_{s^{*},d^{*}})\bar{b}_{\tilde{V}}\}-[\tilde{V}(s,d)-\tilde{V}(s',d')]\bigg]
\end{align*}
Evaluate $\underline{B}$ by:
\[
\underline{B}=\max_{(s,d),(s',d')\in\mathcal{A}}\{|T[\tilde{V}](s,d)-T[\tilde{V}](s',d')-[\tilde{V}(s,d)-\tilde{V}(s',d')]|\}
\]
\item If $\bar{B}-\underline{B}\leq\tau$ then we are done and $\bar{B}$
is the upper bound. If $\bar{B}-\underline{B}>\tau$ then increase
the density of $\mathcal{A}$ and return to step 2 else continue to
step 4.
\item Take as the bound $b(s,d,d')\bar{B}$.
\end{enumerate}
\end{algorithm}

Note that at each iteration in the algorithm above, the upper bound $\bar{B}$ exceeds
$\sup_{s,s',d,d'}|T[\tilde{V}](s,d)-T[\tilde{V}](s',d')-[\tilde{V}(s,d)-\tilde{V}(s',d')]|$.
It is not too difficult to see that $\underline{B}$ lower bounds this quantity. Hence $\tau$ bounds the conservativeness of the
upper bound $\bar{B}$. In Theorem 3 we provide simple conditions
that ensure the algorithm above halts and therefore guarantees we
find a bound $\bar{B}$ that is within $\tau$ of the quantity we
wish to bound.

We define the density of $\mathcal{A}$ as the reciprocal of $\sup_{s\in\mathcal{S},d\in\mathcal{D}}\min_{(s',d)\in\mathcal{A}}||s-s'||$
(if the denominator is zero we take the density to be infinity).

\theoremstyle{plain}
\newtheorem*{Th3}{Theorem 3}
\begin{Th3}
If $\mathcal{S}\subset\mathbb{R}^{k}$ is compact and the functions
$s\mapsto T[\tilde{V}](s,d)$, $s\mapsto\tilde{V}(s,d)$, and $(s,s')\mapsto\delta(F_{s,d},F_{s',d})$
are each uniformly continuous for all $d\in\mathcal{D}$, then as
the density of $\mathcal{A}$ increases to infinity $\bar{B}-\underline{B}\to0$.
\end{Th3}

Theorem 3 implies that the algorithm described above must eventually
halt and thus ensure that $\bar{B}$ exceeds the bound in Theorem 1 by no more than the pre-specified
tolerance $\tau$.

\section{Subset Estimation and Robust Inference}

Let us reintroduce parameters $\theta$, $\beta$, and $F$ into our
notation for the value function $V$ and Bellman operator. For any
bounded function $\tilde{V}$, The results in the previous section
imply an upper-bound of the form:
\begin{eqnarray*}
 &  & \big|\big(\tilde{V}(s,d)-\tilde{V}(s,d')\big)-\Delta_{(\theta,\beta,F)}V(s,d,d')\big)\big|\\
 & \leq & Q_{(\theta,\beta,F)}(s,d,d';\tilde{V})
\end{eqnarray*}

Where $Q_{(\theta,\beta,F)}$ is the bound given in Theorem 2 (or
possibly Theorem 1 if this can be feasibly calculated) for the particular
choice of $\theta$, $\beta$, $F$. Note then that for any $\tilde{V}$:
\begin{eqnarray*}
 &  & \Delta_{(\theta,\beta,F)}\tilde{V}(s;d',d)-Q_{(\theta,\beta,F)}(s,d,d';\tilde{V})\\
 & \leq & \Delta_{(\theta,\beta,F)}V(s;d',d)\\
 & \leq & \Delta_{(\theta,\beta,F)}\tilde{V}(s;d',d)+Q_{(\theta,\beta,F)}(s,d,d';\tilde{V})
\end{eqnarray*}
For any bounded function $\hat{V}:\mathcal{D}\to\mathbb{R}$ let:
\[
L_{(\theta,\beta,F)}(s,d;\hat{V})=P_{\theta}\bigg[\Delta_{d,d'}\epsilon_{t}\geq\Delta_{d',d}u(s;\theta)+\beta\hat{V}(d'),\forall d'\in\mathcal{D},d'\neq d\bigg]
\]

Note that $L_{(\theta,\beta,F)}(s,d)=L_{(\theta,\beta,F)}\big(s,d;\Delta V_{(\theta,\beta,F)}(s,d,\cdot)\big)$.
It is easy to see that $L_{(\theta,\beta,F)}(s,d;\hat{V})$ is monotonically
decreasing in $\hat{V}(d')$ for each $d'\neq d$. As such, for any
bounded $\tilde{V}:\mathcal{S}\times\mathcal{D}\to\mathbb{R}$:
\begin{align*}
 & L_{(\theta,\beta,F)}\big(s,d;\tilde{V}(s,\cdot)-\tilde{V}(s,d)+Q_{(\theta,\beta,F)}(s,d,\cdot;\tilde{V})\big)\\
\leq & L_{(\theta,\beta,F)}(s,d)\\
\leq & L_{(\theta,\beta,F)}\big(s,d;\tilde{V}(s,\cdot)-\tilde{V}(s,d)-Q_{(\theta,\beta,F)}(s,d,\cdot;\tilde{V})\big)
\end{align*}

So we have found upper and lower bounds the likelihood. Thus we can define an upper-
bound on the sample log likelihood of the data by:
\begin{align*}
 & \mathcal{L}_{\{S_{t},D_{t}\}_{t=1}^{T}}^{U}(\theta,\beta;F,\tilde{V})\\
= & \sum_{t=1}^{T}\ln L_{(\theta,\beta,F)}\big(S_{t},D_{t};\tilde{V}(s,\cdot)-\tilde{V}(s,d)-Q_{(\theta,\beta,F)}(S_{t},D_{t},\cdot;\tilde{V})\big)
\end{align*}

And a lower bound by:
\begin{align*}
 & \mathcal{L}_{\{S_{t},D_{t}\}_{t=1}^{T}}^{L}(\theta,\beta;F,\tilde{V})\\
= & \sum_{t=1}^{T}\ln L_{(\theta,\beta,F)}\big(S_{t},D_{t};\tilde{V}(s,\cdot)-\tilde{V}(s,d)+Q_{(\theta,\beta,F)}(S_{t},D_{t},\cdot;\tilde{V})\big)
\end{align*}

Now suppose that $\hat{\theta}^{*}$ and $\hat{\beta}^{*}$ maximize
the exact log likelihood $\mathcal{L}_{\{S_{t},D_{t}\}_{t=1}^{T}}(\theta,\beta;\hat{F})$
(the log-likelihood obtained if the full dynamic programming solution
were available). Suppose we have some approximate dynamic programming
algorithm that, for each $\theta$, $\beta$ and $F$, provides an
approximate value function $\tilde{V}_{(\theta,\beta,F)}$. It follows
that:

\[
\mathcal{L}_{\{S_{t},D_{t}\}_{t=1}^{T}}^{U}(\hat{\theta}^{*},\hat{\beta}^{*};\hat{F},\tilde{V}_{(\hat{\theta}^{*},\hat{\beta}^{*},\hat{F})})\geq\sup_{\theta,\beta}\mathcal{L}_{\{S_{t},D_{t}\}_{t=1}^{T}}^{L}(\theta,\beta;\hat{F},\tilde{V}_{(\theta,\beta,\hat{F})})
\]

Note that the inequality above applies regardless of the algorithm
used to generate $\tilde{V}_{(\theta,\beta,F)}$. The above implies
that $(\hat{\theta}^{*},\hat{\beta}^{*})$ lies in the following set:
\[
\bigg[(\theta,\beta):\mathcal{L}_{\{S_{t},D_{t}\}_{t=1}^{T}}^{U}(\theta,\beta;\hat{F},\tilde{V}_{(\theta,\beta,\hat{F})})\geq\sup_{\theta',\beta'}\mathcal{L}_{\{S_{t},D_{t}\}_{t=1}^{T}}^{L}(\theta',\beta';\hat{F},\tilde{V}_{(\theta',\beta',\hat{F})})\bigg]
\]

The set above can be used as a set estimator for the true parameters
$\theta$ and $\beta$. By construction the set estimator must contain
the infeasible maximum likelihood estimates, that is, the estimates
obtained if the full dynamic programming solution were available.
If the infeasible maximum likelihood estimates $\hat{\theta}^{*}$
and $\hat{\beta}^{*}$ are consistent estimates, then set estimator
is consistent (in the Hausdorff metric).

We can also use the upper and lower bounds on the likelihood to derive
confidence sets that are valid regardless of the accuracy of the approximate
value function $\tilde{V}_{(\theta,\beta,\hat{F})}$. The standard
likelihood-ratio confidence set for the parameters is given by:
\[
[(\theta,\beta):\:2\big(\sup_{\theta',\beta'}\mathcal{L}_{\{S_{t},D_{t}\}_{t=1}^{T}}(\theta',\beta';\hat{F})-\mathcal{L}_{\{S_{t},D_{t}\}_{t=1}^{T}}(\theta,\beta;\hat{F})\big)\leq c_{1-\alpha}]
\]
 where $c_{1-\alpha}$ is the $1-\alpha$-level critical value, usually
the $\alpha$-quantile of the chi-squared distribution with degrees
of freedom equal to the dimension of $\theta$ and $\beta$. The above
is infeasible, it cannot be evaluated because $\mathcal{L}_{\{S_{t},D_{t}\}_{t=1}^{T}}$
incorporates the exact solution to the dynamic programming problem.
However the confidence set above is necessarily a subset of the following
feasible set:
\[
[(\theta,\beta):\:2\big(\sup_{\theta',\beta'}\mathcal{L}_{\{S_{t},D_{t}\}_{t=1}^{T}}^{U}(\theta',\beta';\hat{F},\tilde{V}_{(\theta,\beta,\hat{F})})-\mathcal{L}_{\{S_{t},D_{t}\}_{t=1}^{T}}^{L}(\theta,\beta;\hat{F},\tilde{V}_{(\theta,\beta,\hat{F})})\big)\leq c_{1-\alpha}]
\]

Thus, if the exact, infeasible likelihood ratio-based confidence set
has asymptotically correct size, then so too must the feasible one
based on approximate dynamic programming above.

\section{Monte Carlo Exercise}

As a practical demonstration of our methods we present a Monte Carlo
simulation. We simulate data from a modified Rust model (\cite{Rust1987}).
We apply our methods to account for approximation error in maximum
likelihood estimates that employ approximate dynamic programming.
We evaluate set estimates and robust confidence sets when the approximate
dynamic programming is applied with differing degrees of coarseness.
We compare our set estimates and robust confidence sets to conventional
point estimates and confidence sets that do not account for the error
due to approximation in the value function.

In the modified Rust model, at each period a decision maker chooses
whether or not to send a bus engine for repairs. The choice set is
binary $\mathcal{D}=\{0,1\}$ with $d=0$ representing the decision
not to repair the engine and $d=1$ the decision to repair the engine.
The state space is given by $\mathcal{S}=[0,20]$, each $s\in\mathcal{S}$
represents a possible value for the engine's mileage given in thousands
of miles. We treat the mileage as a metaphor for the general health
the engine and thus it is possible for the mileage to decrease between
periods. The cost of running (and not repairing) an engine increases
with the mileage because with high mileage the engine becomes unreliable
and fuel-hungry. At state $s$ running the engine incurs a cost of
$-\theta_{1}{s}$ where $\theta_{1}$ is a scalar parameter.
Repairing an engine incurs a fixed cost of $-\theta_{2}$ regardless
of the mileage. Thus the deterministic parts of the utilities from
the two choices at state $s$ are given by $u(s,0)=\theta_{1}{s}$
and $u(s,1)=\theta_{2}$.

In each period $t$ there is an additive, stochastic shock to both
the cost of repair and non-repair. The cost shock to non-repair and
repair in period $t$ are respectively $\epsilon_{t}(0)$ and $\epsilon_{t}(1)$.
The shocks are iid and independent between the two choices, they are
observed by the decision-maker before making a decision at time $t$
but are not known prior to $t$. We let the shocks be standard type-1
extreme value distributed. The bus's route is subject to unexpected
changes due to road closures and driver error, as such the mileage
evolves stochastically. If the decision at time $t$ is $D_{t}$ and
the engine has mileage $S_{t}$, then the mileage in period $t+1$
is distributed according to:
\[
S_{t+1}|S_{t},D_{t}\sim\min\{\max\{0,S_{t}+1\{D_{t}=0\}\gamma_{1}+1\{D_{t}=1\}\gamma_{2}+U[-\gamma_{3},\gamma_{3}]\},50\}
\]

Where $U[-\gamma_{3},\gamma_{3}]$ is a uniform random variable between
$-\gamma_{3}$ and $\gamma_{3}$. $\gamma_{1}$, $\gamma_{2}$, and
$\gamma_{3}$ are model parameters. In our simulations $\gamma_{2}$
is negative, thus repair of the engine reduces the mileage.

The modified replacement model differs from Rust's original formulation
in two key ways. Firstly, we treat the mileage as a continuous variable,
In the original model estimated by Rust the mileage is discretized.
Secondly, in the modified model the decision maker chooses whether
or not to repair the engine, reducing its effective mileage by some
fixed amount. In the original model the decision maker chooses whether
or not to replace the engine, bringing the mileage down to zero. The
use of a continuous state-space allows us to examine our method when
the value function is approximated with different degrees of coarseness.
Repairing rather than replacing the engine ensures that the total
variation distance $\delta(F_{s,0},F_{s,1})$ is strictly less than
unity even for large values of $s$. Without this feature the factor
$b(s,d,d')$ in the upper bound in Theorem 1 would be equal to $\frac{1}{1-\beta}$
when the mileage $s$ is large. Thus this modeling decision allows
us to demonstrate the improvement our specific bound provides over
the more crude bound with a factor of $\frac{1}{1-\beta}$ when the
distribution of future states from different choices overlaps.

Note that in the model above:
\[
b(s,d,d')\leq\frac{1-\beta+\beta\min\{|\frac{\gamma_{1}-\gamma_{2}}{2\gamma_{3}}|,1\}}{1-\beta}
\]

Note also that because $\epsilon_{t}(d)$ is independently type-1
extreme value distributed for each $d\in\mathcal{D}$ we have for
any $\tilde{V}$:
\[
T[\tilde{V}](s,d)=E\big[ln\big(\sum_{d\in\mathcal{D}}\exp(u(S_{t+1},d)+\beta\tilde{V}(S_{t+1},d)\big)\big|S_{t}=s,D_{t}=d\big]+c
\]

Where $c$ is the Euler--Mascheroni constant. In order to generate
data from the model we must solve for the value function. An exact
solution is infeasible and so we apply approximate value function
iteration. We employ a dense grid $\tilde{\mathcal{S}}$ that consists
of $1001$ evenly spaced points between $0$ and $20$. We replace
the outer expectation in the RHS above by an empirical expectation
(over $100$ draws of $S_{t+1}$):
\[
T[\tilde{V}](s,d)\approx\frac{1}{n}\sum_{i=1}^{n}ln\big(\sum_{d\in\mathcal{D}}\exp(u(S_{s,d,i},d)+\beta\tilde{V}(S_{s,d,i},d)\big)+c
\]

Where $S_{s,d,i}$ is the $i^{th}$ draw from the distribution of
$S_{t+1}$ given $S_{t}=s$ and $D_{t}=d$. We begin with an initial
guess $V_{0}$ of the value function (our initial guess is that the
function equals zero everywhere), and at the $j^{th}$ iteration we
evaluate $T[V_{j}](s,d)$ for each $s$ in the dense grid $\tilde{\mathcal{S}}$
and each $d\in\mathcal{D}$. We then linearly interpolate the resulting
function between the points in $\tilde{\mathcal{S}}$ in order to
evaluate $V_{j+1}(S_{s,d,i},d)$ for each $d\in\mathcal{D}$, $s\in\tilde{\mathcal{S}}$
and $i\in\{1,...,n\}$. Note that the samples $\{\{\{S_{s,d,i}\}_{i=1}^{n}\}_{d\in\mathcal{D}}\}_{s\in\tilde{\mathcal{S}}}$
are drawn before the first iteration and do not change between iterations.
We continue until the change in the value function between iterations
is smaller than a pre-specified tolerance at all $s\in\tilde{\mathcal{S}}$
and $d\in\mathcal{D}$.

Each simulated dataset consists of $5000$ periods of data generated
from the model above. Table 1 summarizes the data generating process
and states the parameter values used in the simulation.

\begin{table}[H]
\caption{DGP summary and Parameter Values}

\begin{tabular}{cc}
Model Parameters & DGP Summary\tabularnewline
\hline 
\hline 
$\beta=$$0.8$ & $\big(\epsilon_{t}(0),\epsilon_{t}(1)\big)\sim\text{independent type-1 extreme value}$\tabularnewline
$\theta_{1}=-0.6$ & $u(s,0)=\theta_{1}s$\tabularnewline
$\theta_{2}=-4$ & $u(s,1)=\theta_{2}$\tabularnewline
$\gamma_{1}=1$ & $\mathcal{S}=[0,20]$, $\mathcal{D}=\{0,1\}$\tabularnewline
$\gamma_{2}=-1$ & $S_{t+1}|S_{t},D_{t}=0\sim\min\{\max\{S_{t}+\gamma_{1}+U[-\gamma_{3},\gamma_{3}],20\},0\}$\tabularnewline
$\gamma_{3}=5$ & $S_{t+1}|S_{t},D_{t}=1\sim\min\{\max\{S_{t}+\gamma_{2}+U[-\gamma_{3},\gamma_{3}],20\},0\}$\tabularnewline
$T=1000$ & $\tilde{\mathcal{S}}=\{0,\frac{1}{50},\frac{2}{50},...,\frac{999}{50},20\}$\tabularnewline
\end{tabular}
\end{table}

On each simulated dataset we apply maximum likelihood to estimate
the parameters $\theta_{1}$ and $\theta_{2}$. We treat $\beta$,
$\gamma_{1}$, $\gamma_{2}$, and $\gamma_{3}$ as known\@. For each
candidate set of parameter values in the maximization routine we approximately
solve the dynamic programming problem using Bellman iteration on a
grid $\hat{\mathcal{S}}$ of evenly spaced points in the interval
$[0,20]$ and interpolating as described in Section 1. The interpolation
method used is linear interpolation. Let $\tilde{V}$ be an approximate
value function. In order to evaluate the bound on the RHS of the inequality
in Theorem 1 we maximize the following quantity over all $d,d'\in\mathcal{D}$
and all $s,s'\in\tilde{\mathcal{S}}$ where $\tilde{\mathcal{S}}$
is the dense grid used in the data generating process:
\[
|[T[\tilde{V}](s',d')-T[\tilde{V}](s,d)]-[\tilde{V}(s',d')-\tilde{V}(s,d)]]|
\]

We then multiply the above by the factor $\frac{1-\beta+\beta\min\{|\frac{\gamma_{1}-\gamma_{2}}{2\gamma_{3}}|,1\}}{1-\beta}$ to get the bound in Theorem 1.

To examine the coverage of our robust confidence set and set estimator
we numerically maximize $\mathcal{L}_{\{S_{t},D_{t}\}_{t=1}^{T}}^{L}(\theta,\beta;F,\tilde{V}_{(\theta,\beta,F)})$
over $\theta$. The true parameters $(\theta_{1},\theta_{2})$ lie
in the set estimator if:
\[
\sup_{\theta}\mathcal{L}_{\{S_{t},D_{t}\}_{t=1}^{T}}^{L}(\theta,\beta;F,\tilde{V}_{(\theta,\beta,F)})-\mathcal{L}_{\{S_{t},D_{t}\}_{t=1}^{T}}^{U}\big((\theta_{1},\theta_{2}),\beta;F,\tilde{V}_{(\theta,\beta,F)}\big)\leq0
\]

and the true parameters are contained in the robust confidence set
if:
\[
\sup_{\theta}\mathcal{L}_{\{S_{t},D_{t}\}_{t=1}^{T}}^{L}(\theta,\beta;F,\tilde{V}_{(\theta,\beta,F)})-\mathcal{L}_{\{S_{t},D_{t}\}_{t=1}^{T}}^{U}\big((\theta_{1},\theta_{2}),\beta;F,\tilde{V}_{(\theta,\beta,F)}\big)\leq\frac{1}{2}c_{1-\alpha}
\]
Where $c_{1-\alpha}$ is the $1-\alpha$-quantile of the chi-squared
distribution with two degrees of freedom. The true parameters $(\theta_{1},\theta_{2})$
are in the non-robust likelihood-ratio confidence set if:
\[
\sup_{\theta}\mathcal{L}_{\{S_{t},D_{t}\}_{t=1}^{T}}(\theta,\beta;F,\tilde{V}_{(\theta,\beta,F)})-\mathcal{L}_{\{S_{t},D_{t}\}_{t=1}^{T}}\big((\theta_{1},\theta_{2}),\beta;F,\tilde{V}_{(\theta,\beta,F)}\big)\leq\frac{1}{2}c_{1-\alpha}
\]

We use numerical methods to evaluate the supremum on the LHS above.

We apply our methodology with different densities for the grid $\hat{\mathcal{S}}$
that is used in the approximate dynamic programming routine. Since
the grid density controls the coarseness of the approximation, this
allows us to examine how our set estimation and inference procedures
perform as the quality of the approximation changes.

Table 1 provides results from our simulations. Each column corresponds
to different choice for $\hat{\mathcal{S}}$, the grid used to perform
approximate dynamic programming in estimation. The figures in the
first row are the mean (over $500$ simulations) squared errors of
the maximum likelihood point estimates. In the second row are the
frequencies with which the true parameters lie within the set estimate.
In the third row are the frequencies with which the robust confidence
sets contain the true parameters. Finally, in the fourth row are the
frequencies with which the standard likelihood ratio-based confidence
sets contain the true parameters.

\begin{table}[H]
\caption{Simulation Results}

\begin{tabular}{cccc}
 & \multicolumn{3}{c}{Number of points in $\hat{\mathcal{S}}$:}\tabularnewline
 & $10$ & $100$ & $500$\tabularnewline
\hline 
Mean Squared Error & 0.023 & 0.023 & 0.022\tabularnewline
Set Estimator Coverage & 1 & 1 & 0.908\tabularnewline
Robust Confidence Set Coverage & 1 & 1 & 0.996\tabularnewline
Standard Confidence Set Coverage & 0.702 & 0.942 & 0.938\tabularnewline
\end{tabular}
\end{table}

We see that the standard non-robust confidence set has coverage much
lower than $95\%$ when $\hat{\mathcal{S}}$ contains only $10$ points.
This is not surprising, when the grid used for the approximate dynamic
programming is sparse, the quality of approximation is low leading
to biased point estimates and confidence sets with low coverage. When
the number of points in $\hat{\mathcal{S}}$ is larger the coverage
is close to the stated level of $95\%$. By contrast, the robust confidence set
has coverage of at least $95\%$ regardless of the density of the
grid used in the approximate dynamic programming. The robust confidence
sets are based on worst-case scenarios for the approximation error
and as such they are generally conservative, having coverage greater
than the stated level. As the error from approximation is reduced
and statistical noise dominates, the robust confidence sets should
have coverage approaching the stated $95\%$ level, and indeed we
do see that the coverage is slightly lower in the case of $\hat{\mathcal{S}}$
with $500$ points.

The set estimator is not required to cover the true parameters with
any particular frequency, but is nonetheless interesting to observe
its coverage in our simulations. We note that the set estimates cover
the true parameters in $100\%$ of our simulations when $\hat{\mathcal{S}}$
has either $10$ or $100$ points. When the grid is made dense the
set estimator should shrink around the maximum likelihood point estimates
and the coverage should fall, and indeed when $\hat{\mathcal{S}}$
contains $500$ points the coverage of the set estimator is reduced
to $91\%$.

To get a sense of what the set estimates and confidence sets look
like, for each choice of $\hat{\mathcal{S}}$ we plot the sets for
one simulation draw. We evaluate $\mathcal{L}_{\{S_{t},D_{t}\}_{t=1}^{T}}^{U}$
on a grid of values for $\theta_{1}$ and $\theta_{2}$. We then interpolate
the function over a denser grid and for each set of parameters in
this grid we compare $\mathcal{L}_{\{S_{t},D_{t}\}_{t=1}^{T}}^{U}$
and $\sup_{\theta}\mathcal{L}_{\{S_{t},D_{t}\}_{t=1}^{T}}^{L}(\theta,\beta;F,\tilde{V}_{(\theta,\beta,F)})$
to determine whether the parameters tlie inside the set estimate and
robust confidence set. Similarly, we evaluate $\mathcal{L}_{\{S_{t},D_{t}\}_{t=1}^{T}}$
for each pair of parameter values on the same grid and interpolate
to assess which points lie inside the standard non-robust confidence
set.

We plot these sets in Figure 1 below. Also indicated in each sub-figure
are the maximum likelihood point estimates and the true values of
parameters $\theta_{1}$ and $\theta_{2}$. As one might expect, as
$\hat{\mathcal{S}}$ grows dense, the robust confidence set shrinks
and becomes close to the standard confidence set, and the set estimate
shrinks around the MLE estimator.

In the case of $\hat{\mathcal{S}}$
with $10$ points, the approximation is weak and there is substantial
undercoverage of the non-robust confidence set. We see from Sub-Figure
1.a that in this case the set estimates and robust confidence sets
are very large, reflecting the poor strength of approximation. Note
that because the worst-case approximation error dominates the statistical
noise in this case, the robust confidence set and the set estimates
are visibly indistinguishable. In the case of $\hat{\mathcal{S}}$
with $100$ points the approximation quality is improved and we see
from Sub-Figure 1.b that the set estimates and confidence sets are
much smaller, although still significantly larger than the non-robust
confidence set, which (according to Table 1) has nearly correct coverage
in this case. When $\hat{\mathcal{S}}$ contains $500$ points the
approximation is good and we see from Sub-Figure 1.c that the robust
confidence set is only a little larger than the non-robust, moreover
in this case the set estimate is contained within the non-robust confidence
set.

An empirical researcher using our methodology would see that in the
case of $\hat{\mathcal{S}}$ with $10$ points, the robust confidence
set and set estimate are impractically large, and should conclude
that the quality of approximation is poor. The researcher can then
decide to increase the density of the grid in order to achieve a better
approximation. Thus our methodology allows the researcher
to make a more informed decision about the approximation method and
to better balance the trade-off between computational expedience and
the bias due to approximate dynamic programming.

\begin{figure}[h]
\caption{Simulated Set Estimates}
\subfloat[]{

\includegraphics[scale=0.25]{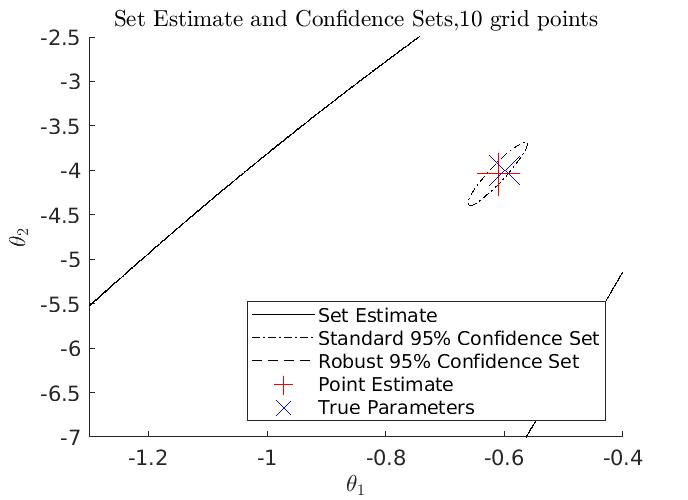}}\subfloat[]{\includegraphics[scale=0.25]{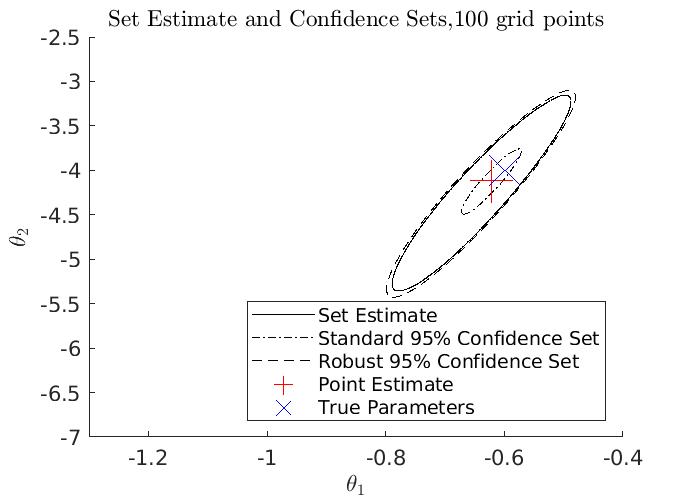}}\\
\subfloat[]{

\includegraphics[scale=0.25]{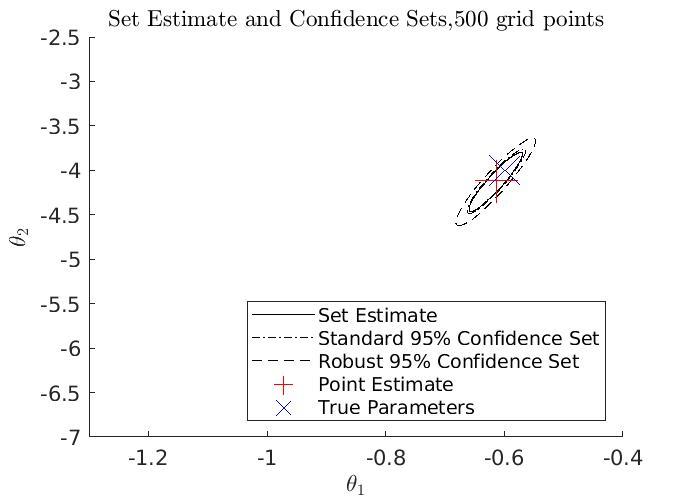}

}
\end{figure}

\section{Conclusion}

We develop an approach to estimation and inference in dynamic discrete
choice models that directly accounts for error due to the use of approximate
dynamic programming. We prove the validity of our approach and provide
simulation evidence that our methodology is of practical use. Of course,
our simulation results pertain to one particular model, and it is
possible that the robust confidence sets are overly conservative in
other settings. To better assess the practical efficacy of our methods
we hope to implement the procedure on real data in future work.

The methodology detailed in the paper is very flexible in that it
applies regardless of the approximation method used. It may be possible
to refine our approach taking into account features of the specific
approximate dynamic programming method used. It may also be possible
to tighten our error bounds by incorporating information about the
value function that can be derived analytically, for example, in some
cases one can prove the value function is convex. This avenue may
also be worth exploring in future research.

\bibliographystyle{authordate1}
\bibliography{approxcites}

\appendix
\section*{Appendix: Proofs}

\begin{proof}[Proof Theorem 1]
By the definition of $T$ for any bounded functions $\tilde{V}$ and
$\hat{V}$ that map from $\mathcal{S}\times\mathcal{D}$ to $\mathbb{R}$:

\begin{eqnarray*}
 & = & T[\tilde{V}](s,d)-T[\tilde{V}](s',d')-\big(T[\hat{V}](s',d')-T[\hat{V}](s',d')\big)\\
 & = & E[E[\max_{d''\in\mathcal{D}}\,u(S_{t+1},d'')+\beta\tilde{V}(S_{t+1},d'')+\epsilon_{t+1}(d'')]|S_{t}=s,D_{t}=d]\\
 & - & E[E[\max_{d''\in\mathcal{D}}\,u(s',d'')+\beta\tilde{V}(s',d'')+\epsilon_{t+1}(d'')]|S_{t}=s',D_{t}=d']\\
 & + & E[E[\max_{d''\in\mathcal{D}}\,u(S_{t+1},d'')+\beta\hat{V}(S_{t+1},d'')+\epsilon_{t+1}(d'')]|S_{t}=s,D_{t}=d]\\
 & - & E[E[\max_{d''\in\mathcal{D}}\,u(S_{t+1},d'')+\beta\hat{V}(S_{t+1},d'')+\epsilon_{t+1}(d'')]|S_{t}=s',D_{t}=d']
\end{eqnarray*}

Because $\epsilon_{t+1}$ and $S_{t+1}$ are independent we can rewrite
the right-hand side of the final equality above as an integral:

\begin{eqnarray*}
 &  & \int\bigg(E[\max_{d''\in\mathcal{D}}\,u(s'',d'')+\beta\tilde{V}(s'',d'')+\epsilon_{t}(d'')]\\
 &  & -E[\max_{d''\in\mathcal{D}}\,u(s'',d'')+\beta\hat{V}(s'',d'')+\epsilon_{t}(d'')]\bigg)d(F_{s,d}-F_{s',d'})
\end{eqnarray*}

It is an elementary property of the total variation distance that
for any bounded function $g:\mathcal{S}\to\mathbb{R}$:
\begin{eqnarray*}
\int g(s'')d(F_{s,d}-F_{s',d'}) & \leq & \delta(F_{s,d},F_{s',d'})\big(\sup_{s\in\mathcal{S}}g(s)-\inf_{s\in\mathcal{S}}g(s)\big)
\end{eqnarray*}

Since $u$, $\tilde{V}$, and $\hat{V}$ are bounded and $E[\max_{d\in\mathcal{D}}\epsilon_{t}(d)]<\infty$,
$E[\max_{d\in\mathcal{D}}\,u(s,d)+\beta\tilde{V}(s,d)+\epsilon_{t}(d)]$
and $E[\max_{d\in\mathcal{D}}\,u(s,d)+\beta\hat{V}(s,d)+\epsilon_{t}(d)]$
are also uniformly bounded over $s\in\mathcal{S}$ and so applying
the inequality above:

\begin{eqnarray*}
 &  & \int\bigg(E[\max_{d''\in\mathcal{D}}\,u(s'',d'')+\beta\tilde{V}(s'',d'')+\epsilon_{t}(d'')]\\
 &  & -E[\max_{d''\in\mathcal{D}}\,u(s'',d'')+\beta\hat{V}(s'',d'')+\epsilon_{t}(d'')]\bigg)d(F_{s,d}-F_{s',d'})\\
 & \leq & \delta(F_{s,d},F_{s',d'})\sup_{s''\in\mathcal{S}}\bigg(E[\max_{d''\in\mathcal{D}}\,u(s'',d'')+\beta\tilde{V}(s'',d'')+\epsilon_{t}(d'')]\\
 &  & -E[\max_{d''\in\mathcal{D}}\,u(s'',d'')+\beta\hat{V}(s'',d'')+\epsilon_{t}(d'')]\bigg)\\
 & - & \delta(F_{s,d},F_{s',d'})\inf_{s''\in\mathcal{S}}\bigg(E[\max_{d''\in\mathcal{D}}\,u(s'',d'')+\beta\tilde{V}(s'',d'')+\epsilon_{t}(d'')]\\
 &  & -E[\max_{d''\in\mathcal{D}}\,u(s'',d'')+\beta\hat{V}(s'',d'')+\epsilon_{t}(d'')]\bigg)
\end{eqnarray*}

It is easy to see that:
\begin{align*}
 & \sup_{s\in\mathcal{S}}\bigg(E[\max_{d\in\mathcal{D}}\,u(s,d)+\beta\tilde{V}(s,d)+\epsilon_{t}(d)]\\
 & -E[\max_{d\in\mathcal{D}}\,u(s,d)+\beta\hat{V}(s,d)+\epsilon_{t}(d)]\bigg)\\
\leq & \beta\sup_{s\in\mathcal{S}}\max_{d\in\mathcal{D}}\big(\tilde{V}(s,d)-\hat{V}(s,d)\big)
\end{align*}

And:
\begin{align*}
 & \inf_{s\in\mathcal{S}}\bigg(E[\max_{d}\,u(s,d)+\beta\tilde{V}(s,d)+\epsilon_{t}(d)]\\
 & -E[\max_{d''}\,u(s,d)+\beta\hat{V}(s,d)+\epsilon_{t}(d)]\bigg)\\
\geq & \beta\inf_{s}\min_{d\in\mathcal{D}}\big(\tilde{V}(s,d)-\hat{V}(s,d)\big)
\end{align*}

Combining we get:
\begin{eqnarray*}
 &  & T[\tilde{V}](s,d)-T[\tilde{V}](s',d')-\big(T[\hat{V}](s',d')-T[\hat{V}](s',d')\big)\\
 & \leq & \delta(F_{s,d},F_{s',d'})\beta\bigg(\sup_{s\in\mathcal{S}}\max_{d\in\mathcal{D}}\big(\tilde{V}(s,d)-\hat{V}(s,d)\big)-\inf_{s}\min_{d\in\mathcal{D}}\big(\tilde{V}(s,d)-\hat{V}(s,d)\big)\bigg)\\
 & = & \delta(F_{s,d},F_{s',d'})\beta\sup_{s,s'\in\mathcal{S},d,d'\in\mathcal{D}}\big|\tilde{V}(s,d)-\tilde{V}(s',d')-\big(\hat{V}(s,d)-\hat{V}(s',d')\big)\big|
\end{eqnarray*}

Since the same reasoning holds with $\tilde{V}$ and $\hat{V}$ switched
we get:
\begin{eqnarray*}
 &  & \big|T[\tilde{V}](s,d)-T[\tilde{V}](s',d')-\big(T[\hat{V}](s',d')-T[\hat{V}](s',d')\big)\big|\\
 & \leq & \delta(F_{s,d},F_{s',d'})\beta\sup_{s,s'\in\mathcal{S},d,d'\in\mathcal{D}}\big|\tilde{V}(s,d)-\tilde{V}(s',d')-\big(\hat{V}(s,d)-\hat{V}(s',d')\big)\big|
\end{eqnarray*}

Now, $V$ is a fixed point of $T$ and it must be bounded because
$u$ is bounded and $E[\max_{d\in\mathcal{D}}\epsilon_{t}(d)]<\infty$.
Thus the above implies:
\begin{eqnarray}
 &  & \big|T[\tilde{V}](s,d)-T[\tilde{V}](s',d')-\big(V(s',d')-V(s',d')\big)\big|\nonumber \\
 & \leq & \delta(F_{s,d},F_{s',d'})\beta\sup_{s,s'\in\mathcal{S},d,d'\in\mathcal{D}}\big|\tilde{V}(s,d)-\tilde{V}(s',d')-\big(V(s,d)-V(s',d')\big)\big|\label{eq:stpp}
\end{eqnarray}

Taking the supremum over both sides:

\begin{eqnarray*}
 &  & \sup_{s,s'\in\mathcal{S},d,d'\in\mathcal{D}}\big|T[\tilde{V}](s,d)-T[\tilde{V}](s',d')-\big(V(s',d')-V(s',d')\big)\big|\\
 & \leq & \sup_{s,s'\in\mathcal{S},d,d'\in\mathcal{D}}\delta(F_{s,d},F_{s',d'})\beta\sup_{s,s'\in\mathcal{S},d,d'\in\mathcal{D}}\big|\tilde{V}(s,d)-\tilde{V}(s',d')-\big(V(s,d)-V(s',d')\big)\big|
\end{eqnarray*}

By the reverse triangle inequality:
\begin{eqnarray*}
 &  & \sup_{s,s'\in\mathcal{S},d,d'\in\mathcal{D}}\big|\tilde{V}(s,d)-\tilde{V}(s',d')-\big(V(s',d')-V(s',d')\big)\big|\\
 & - & \sup_{s,s'\in\mathcal{S},d,d'\in\mathcal{D}}\big|T[\tilde{V}](s,d)-T[\tilde{V}](s',d')-\big(\tilde{V}(s',d')-\tilde{V}(s',d')\big)\big|\\
 & \leq & \sup_{s,s'\in\mathcal{S},d,d'\in\mathcal{D}}\delta(F_{s,d},F_{s',d'})\beta\sup_{s,s'\in\mathcal{S},d,d'\in\mathcal{D}}\big|\tilde{V}(s,d)-\tilde{V}(s',d')-\big(V(s,d)-V(s',d')\big)\big|
\end{eqnarray*}

Since $\delta(F_{s,d},F_{s',d'})\beta\leq\beta<1$ we can solve to
get:

\begin{eqnarray}
 &  & \sup_{s,s'\in\mathcal{S},d,d'\in\mathcal{D}}\big|\tilde{V}(s,d)-\tilde{V}(s',d')-\big(V(s',d')-V(s',d')\big)\big|\nonumber \\
 & \leq & \frac{1}{1-\sup_{s,s'\in\mathcal{S},d,d'\in\mathcal{D}}\delta(F_{s,d},F_{s',d'})\beta}\nonumber \\
 & \times & \sup_{s,s'\in\mathcal{S},d,d'\in\mathcal{D}}\big|T[\tilde{V}](s,d)-T[\tilde{V}](s',d')-\big(\tilde{V}(s',d')-\tilde{V}(s',d')\big)\big|\label{eq:stpp2}
\end{eqnarray}

Also by the triangle inequality:
\begin{eqnarray*}
 &  & |\tilde{V}(s,d)-\tilde{V}(s',d')-[V(s,d)-V(s',d')]|\\
 & \leq & |T[\tilde{V}](s,d)-T[\tilde{V}](s',d')-[\tilde{V}(s,d)-\tilde{V}(s',d')]|\\
 & + & |T[\tilde{V}](s,d)-T[\tilde{V}](s',d')-[V(s,d)-V(s',d')]|
\end{eqnarray*}

Substituting using (\ref{eq:stpp}) we get:
\begin{eqnarray*}
 &  & |\tilde{V}(s,d)-\tilde{V}(s',d')-[V(s,d)-V(s',d')]|\\
 & \leq & |T[\tilde{V}](s,d)-T[\tilde{V}](s',d')-[\tilde{V}(s,d')-\tilde{V}(s',d')]|\\
 & + & \delta(F_{s,d},F_{s',d'})\beta\sup_{s,s'\in\mathcal{S},d,d'\in\mathcal{D}}\big|\tilde{V}(s,d)-\tilde{V}(s',d')-\big(V(s,d)-V(s',d')\big)\big|
\end{eqnarray*}

And so from (\ref{eq:stpp2}):
\begin{eqnarray*}
 &  & |\tilde{V}(s,d)-\tilde{V}(s',d')-[V(s,d)-V(s',d')]|\\
 & \leq & |T[\tilde{V}](s,d)-T[\tilde{V}](s',d')-[\tilde{V}(s,d')-\tilde{V}(s',d')]|\\
 & + & \bigg(\frac{\delta(F_{s,d},F_{s',d'})\beta}{1-\sup_{s,s'\in\mathcal{S},d,d'\in\mathcal{D}}\delta(F_{s,d},F_{s',d'})\beta}\\
 & \times & \sup_{s,s'\in\mathcal{S},d,d'\in\mathcal{D}}\big|T[\tilde{V}](s,d)-T[\tilde{V}](s',d')-\big(\tilde{V}(s',d')-\tilde{V}(s',d')\big)\big|\bigg)
\end{eqnarray*}

Taking the supremum over the RHS we get:

\begin{eqnarray*}
 &  & |\tilde{V}(s,d)-\tilde{V}(s',d')-[V(s,d)-V(s',d')]|\\
 & \leq & \bigg(\frac{1-\sup_{s,s'\in\mathcal{S},d,d'\in\mathcal{D}}\delta(F_{s,d},F_{s',d'})\beta+\delta(F_{s,d},F_{s',d'})\beta}{1-\sup_{s,s'\in\mathcal{S},d,d'\in\mathcal{D}}\delta(F_{s,d},F_{s',d'})\beta}\\
 & \times & \sup_{s,s'\in\mathcal{S},d,d'\in\mathcal{D}}\big|T[\tilde{V}](s,d)-T[\tilde{V}](s',d')-\big(\tilde{V}(s',d')-\tilde{V}(s',d')\big)\big|\bigg)
\end{eqnarray*}

The conclusion then follows immediately.
\end{proof}

\begin{proof}[Proof Theorem 2]
First we show that for any $s,s'\in\mathcal{S}$ and $d,d'\in\mathcal{D}$:

\[
|T\tilde{V}(s,d)-T\tilde{V}(s',d')|\leq\delta(F_{s,d},F_{s',d'})\bar{b}_{\tilde{V}}
\]

By the definition of $T$:

\begin{eqnarray*}
 &  & T[\tilde{V}](s,d)-T[\tilde{V}](s',d')\\
 & = & E[E[\max_{d''\in\mathcal{D}}\,u(S_{t+1},d'')+\beta\tilde{V}(S_{t+1},d'')+\epsilon_{t+1}(d'')]|S_{t}=s,D_{t}=d]\\
 & - & E[E[\max_{d''\in\mathcal{D}}\,u(s',d'')+\beta\tilde{V}(s',d'')+\epsilon_{t+1}(d'')]|S_{t}=s',D_{t}=d']
\end{eqnarray*}

Using independence of $\epsilon_{t+1}$ and $S_{t+1}$ the right-hand
side above is equal to:

\begin{eqnarray*}
 &  & \int E[\max_{d''\in\mathcal{D}}\,u(s'',d'')+\beta\tilde{V}(s'',d'')+\epsilon_{t}(d'')]d(F_{s,d}-F_{s',d'})
\end{eqnarray*}

As in the proof of Theorem 1 we can apply elementary properties of
the total variation distance to get:
\[
\int g(x)d(F_{s,d}-F_{s',d'})\leq\delta(F_{s,d},F_{s',d'})\sup_{x,x'\in\mathcal{S}(s,d)\cup\mathcal{S}(s',d')}g(x)-g(x')
\]
\begin{eqnarray*}
 &  & \int E[\max_{d''\in\mathcal{D}}\,u(s'',d'')+\beta\tilde{V}(s'',d'')+\epsilon_{t}(d'')]d(F_{s,d}-F_{s',d'})\\
 & \leq & \delta(F_{s,d},F_{s',d'})\bigg(\sup_{s'\in\mathcal{S}}E[\max_{d''\in\mathcal{D}}\,u(s'',d'')+\beta\tilde{V}(s'',d'')+\epsilon_{t}(d'')]\\
 &  & -\inf_{s'\in\mathcal{S}}E[\max_{d''\in\mathcal{D}}\,u(s'',d'')+\beta\tilde{V}(s'',d'')+\epsilon_{t}(d'')]\bigg)\\
 & = & \delta(F_{s,d},F_{s',d'})\bigg(\sup_{s',s\in\mathcal{S}}E[\max_{d''\in\mathcal{D}}\,u(s'',d'')+\beta\tilde{V}(s'',d'')+\epsilon_{t}(d'')]\\
 &  & -E[\max_{d''\in\mathcal{D}}\,u(s'',d'')+\beta\tilde{V}(s'',d'')+\epsilon_{t}(d'')]\bigg)
\end{eqnarray*}

It is easy to see that for any $s'\in\mathcal{S}$:

\begin{align*}
 & E[\max_{d\in\mathcal{D}}\,u(s,d)+\beta\tilde{V}(s,d)+\epsilon_{t}(d)]-E[\max_{d'\in\mathcal{D}}\,u(s',d)+\beta\tilde{V}(s',d)+\epsilon_{t}(d)]\bigg)\\
\leq & \max_{d\in\mathcal{D}}\,\big(u(s,d)+\beta\tilde{V}(s,d)-u(s',d)+\beta\tilde{V}(s',d)\big)
\end{align*}

And so in all:
\begin{align*}
 & T[\tilde{V}](s,d)-T[\tilde{V}](s',d')\\
\leq & \delta(F_{s,d},F_{s',d'})\sup_{s,s'\in\mathcal{S},d\in\mathcal{D},}\,\big(u(s,d)+\beta\tilde{V}(s,d)-u(s',d)+\beta\tilde{V}(s',d)\big)\\
= & \delta(F_{s,d},F_{s',d'})\big|\sup_{s,s'\in\mathcal{S},d\in\mathcal{D},}\,\big(u(s,d)+\beta\tilde{V}(s,d)-u(s',d)+\beta\tilde{V}(s',d)\big)\big|
\end{align*}

By symmetry:
\begin{align*}
 & |T[\tilde{V}](s,d)-T[\tilde{V}](s',d')|\\
\leq & \delta(F_{s,d},F_{s',d'})\big|\sup_{s,s'\in\mathcal{S},d\in\mathcal{D},}\,\big(u(s,d)+\beta\tilde{V}(s,d)-u(s',d)+\beta\tilde{V}(s',d)\big)\big|\\
= & \delta(F_{s,d},F_{s',d'})\bar{b}_{\tilde{V}}
\end{align*}

Now adding and subtracting terms, for any $(s^{*},d^{*}),(s^{**},d^{**})\in\mathcal{A}$:
\begin{eqnarray*}
 &  & T[\tilde{V}](s,d)-T[\tilde{V}](s',d')-[\tilde{V}(s,d)-\tilde{V}(s',d')]\\
 & = & T[\tilde{V}](s^{*},d^{*})+T[\tilde{V}](s,d)-T[\tilde{V}](s^{*},d^{*})\\
 & - & T[\tilde{V}](s^{**},d^{**})+T[\tilde{V}](s^{**},d^{**})-T[\tilde{V}](s',d')\\
 & - & [\tilde{V}(s,d)-\tilde{V}(s',d')]\\
 & \leq & T[\tilde{V}](s^{*},d^{*})+|T[\tilde{V}](s,d)-T[\tilde{V}](s^{*},d^{*})|\\
 & - & T[\tilde{V}](s^{**},d^{**})+|T[\tilde{V}](s',d')-T[\tilde{V}](s^{**},d^{**})|\\
 & - & [\tilde{V}(s,d)-\tilde{V}(s',d')]
\end{eqnarray*}

Using the bound we derived earlier we get from the above:
\begin{eqnarray*}
 &  & T[\tilde{V}](s,d)-T[\tilde{V}](s',d')-[\tilde{V}(s,d)-\tilde{V}(s',d')]\\
 & \leq & T[\tilde{V}](s^{*},d^{*})+\delta(F_{s,d},F_{s^{*},d^{*}})\bar{b}_{\tilde{V}}\\
 & - & T[\tilde{V}](s^{**},d^{**})+\delta(F_{s',d'},F_{s^{**},d^{**}})\bar{b}_{\tilde{V}}\\
 & - & [\tilde{V}(s,d)-\tilde{V}(s',d')]
\end{eqnarray*}

Since the above holds for any $(s^{*},d^{*}),(s^{**},d^{**})\in\mathcal{A}$
we can choose these to minimize it:
\begin{eqnarray*}
 &  & T[\tilde{V}](s,d)-T[\tilde{V}](s',d')-[\tilde{V}(s,d)-\tilde{V}(s',d')]\\
 & \leq & \min_{(s^{*},d^{*})\in\mathcal{A}}\{T[\tilde{V}](s^{*},d^{*})+\delta(F_{s,d},F_{s^{*},d^{*}})\bar{b}_{\tilde{V}}\}\\
 & - & \max_{(s^{*},d^{*})\in\mathcal{A}}\{T[\tilde{V}](s^{*},d^{*})-\delta(F_{s',d'},F_{s^{*},d^{*}})\bar{b}_{\tilde{V}}\}\\
 & - & [\tilde{V}(s,d)-\tilde{V}(s',d')]
\end{eqnarray*}

Taking the supremum of both sides above over $s,s'\in\mathcal{S}$
and $d,d'\in\mathcal{D}$ and noting that by symmetry:

\[
\sup_{s,s'\in\mathcal{S},d,d'\in\mathcal{D}}T[\tilde{V}](s,d)-T[\tilde{V}](s',d')-[\tilde{V}(s,d)-\tilde{V}(s',d')]>0
\]

We get the result.
\end{proof}

\begin{proof}[Proof Theorem 3]
Let $B$ be defined by:
\[
B=\sup_{s,s'\in S,d,d'\in D}|[T[\tilde{V}](s',d')-T[\tilde{V}](s,d)]-[\tilde{V}(s',d')-\tilde{V}(s,d)]]|
\]

Fix some $\tau>0$. Recall $\underline{B}$ is defined by:
\begin{align*}
\underline{B} & =\max_{(s,d),(s',d')\in\mathcal{A}}\{|T[\tilde{V}](s,d)-T[\tilde{V}](s',d')-[\tilde{V}(s,d)-\tilde{V}(s',d')]|\}\\
 & =\max_{(s,d)\in\mathcal{A}}\big(T[\tilde{V}](s,d)-\tilde{V}(s,d)\big)-\min_{(s',d')\in\mathcal{A}}\big(T[\tilde{V}](s',d')-\tilde{V}(s',d')\big)
\end{align*}

A linear combination of two uniformly continuous functions is also
uniformly continuous, and so $s\mapsto T[\tilde{V}](s,d)-\tilde{V}(s,d)$
is uniformly continuous for each $d\in\mathcal{D}$, there must exist
some $\eta_{1}>0$ so that for every $d\in\mathcal{D}$:
\[
\sup_{s,s'\in\mathcal{S}:||s-s'||\leq\eta}|\big(T[\tilde{V}](s,d)-\tilde{V}(s,d)\big)-\big(T[\tilde{V}](s',d)-\tilde{V}(s',d)\big)|\leq\frac{1}{4}\tau
\]

So suppose that for any $d\in\mathcal{D}$ and $s\in\mathcal{S}$
there is an $s'$ with $||s-s'||\leq\eta_{1}$ and $(s',d)\in\mathcal{A}$,
then:
\[
\sup_{s\in\mathcal{S},d\in\mathcal{D}}\min_{s'\in\mathcal{S}:(s',d)\in\mathcal{A}}|\big(T[\tilde{V}](s,d)-\tilde{V}(s,d)\big)-\big(T[\tilde{V}](s',d)-\tilde{V}(s',d)\big)|\leq\frac{1}{4}\tau
\]

And so:
\[
\sup_{s\in\mathcal{S},d\in\mathcal{D}}\big(T[\tilde{V}](s,d)-\tilde{V}(s,d)\big)-\max_{(s,d)\in\mathcal{A}}\big(T[\tilde{V}](s,d)-\tilde{V}(s,d)\big)\leq\frac{1}{4}\tau
\]

And:
\[
\inf_{s\in\mathcal{S},d\in\mathcal{D}}\big(T[\tilde{V}](s,d)-\tilde{V}(s,d)\big)-\min_{(s,d)\in\mathcal{A}}\big(T[\tilde{V}](s,d)-\tilde{V}(s,d)\big)\leq\frac{1}{4}\tau
\]

And so:
\[
B-\underline{B}\leq\frac{1}{2}\tau
\]

Next, recall the definition of $\bar{B}$:

\begin{align*}
\bar{B}= & \sup_{s,s'\in S,d,d'\in D}\bigg[\min_{(s^{*},d^{*})\in\mathcal{A}}\{T[\tilde{V}](s^{*},d^{*})+\delta(F_{s,d},F_{s^{*},d^{*}})\bar{b}_{\tilde{V}}\}\\
 & -\max_{(s^{*},d^{*})\in\mathcal{A}}\{T[\tilde{V}](s^{*},d^{*})-\delta(F_{s',d'},F_{s^{*},d^{*}})\bar{b}_{\tilde{V}}\}-[\tilde{V}(s,d)-\tilde{V}(s',d')]\bigg]\\
= & \sup_{s,s'\in S,d,d'\in D}\bigg[\min_{(s^{*},d^{*})\in\mathcal{A}}\big\{ T[\tilde{V}](s^{*},d^{*})-T[\tilde{V}](s,d)\\
 & +\big(\delta(F_{s,d},F_{s^{*},d^{*}})-\delta(F_{s,d},F_{s,d})\big)\bar{b}_{\tilde{V}}\big\}\\
 & +\min_{(s^{*},d^{*})\in\mathcal{A}}\big\{ T[\tilde{V}](s',d')-T[\tilde{V}](s^{*},d^{*})\\
 & +\big(\delta(F_{s',d'},F_{s^{*},d^{*}})-\delta(F_{s',d'},F_{s',d'})\big)\bar{b}_{\tilde{V}}\big\}\\
 & -[\big(\tilde{V}(s,d)-T[\tilde{V}](s,d)\big)+\big(T[\tilde{V}](s',d')-\tilde{V}(s',d')\big)]\bigg]
\end{align*}

Where we have used that $\delta(F_{s,d},F_{s,d})=\delta(F_{s',d'},F_{s',d'})=0$.
By the triangle inequality, the RHS of the final equality above is
bounded by:
\begin{align*}
 & 2\sup_{s\in S,d\in D}\min_{(s^{*},d^{*})\in\mathcal{A}}|T[\tilde{V}](s^{*},d^{*})-T[\tilde{V}](s,d)+\big(\delta(F_{s,d},F_{s^{*},d^{*}})-\delta(F_{s,d},F_{s,d})\big)\bar{b}_{\tilde{V}}|\\
+ & B
\end{align*}

A linear combination of two uniformly continuous functions is uniformly
continuous and so $(s,s')\mapsto T[\tilde{V}](s,d)+\delta(F_{s',d},F_{s,d})\bar{b}_{\tilde{V}}$
is uniformly continuous for all $d\in\mathcal{D}$. It follows that
for some $\eta_{2}>0$:
\[
\sup_{s,s'\in S:||s-s'||\leq\eta}|T[\tilde{V}](s',d)-T[\tilde{V}](s,d)+\big(\delta(F_{s,d},F_{s',d})-\delta(F_{s,d},F_{s,d})\big)\bar{b}_{\tilde{V}}|\leq\frac{1}{4}\tau
\]

For all $d\in\mathcal{D}$. So suppose that for any $d\in\mathcal{D}$
and $s\in\mathcal{S}$ there is an $s'$ with $||s-s'||\leq\eta_{2}$
and $(s',d)\in\mathcal{A}$, then:

\[
\sup_{s\in S,d\in D}\min_{(s^{*},d^{*})\in\mathcal{A}}|T[\tilde{V}](s^{*},d^{*})-T[\tilde{V}](s,d)+\big(\delta(F_{s,d},F_{s^{*},d^{*}})-\delta(F_{s,d},F_{s,d})\big)\bar{b}_{\tilde{V}}|\leq\frac{1}{2}\tau
\]

And so:
\[
\bar{B}-B\leq\frac{1}{2}\tau
\]

Finally, for dense enough $\mathcal{A}$, for any $d\in\mathcal{D}$
and $s\in\mathcal{S}$ there is an $s'$ with $||s-s'||\leq\min\{\eta_{1},\eta_{2}\}$
and $(s',d)\in\mathcal{A}$ and hence:
\[
\bar{B}-\underline{B}\leq\tau
\]

$\tau$ was chosen arbitrarily and so as the density of $\mathcal{A}$
grows to infinity $\bar{B}-\underline{B}\to0$.
\end{proof}
\end{document}